\documentclass[preprintnumbers,article,amsmath,amssymb,floatfix,10pt,prd,superscriptaddress,nofootinbib,twocolumn]{revtex4}

\usepackage{doi}%<----------

\usepackage{graphicx}% Include figure files
\usepackage{dcolumn}% Align table columns on decimal point
\usepackage{bm}% bold math
\usepackage{color}
\usepackage{enumitem}
\usepackage{amsmath}
\usepackage{amssymb}

\newcommand{\beq}{\begin{equation}}
\newcommand{\eeq}{\end{equation}}
\newcommand{\bea}{\begin{eqnarray}}
\newcommand{\eea}{\end{eqnarray}}

\usepackage{bbm}
\usepackage{amsfonts}
\usepackage{mathrsfs}
\usepackage{latexsym}
\usepackage{epsfig}
\usepackage{epstopdf}
\usepackage{epstopdf}
\usepackage{graphicx}
\usepackage{amssymb}
\usepackage{amsmath}
\usepackage{dcolumn}
\usepackage{bm}
\usepackage{color}
\usepackage{comment}
\usepackage{xcolor}
\usepackage{ulem}
\usepackage{hyperref}
\hypersetup{
  colorlinks=true,        % false: boxed links; true: colored links
  linkcolor=magenta,         % color of internal links
  citecolor=red,         % color of links to bibliography
}

\begin{document}

\title{Gravitational lensing around a Kerr-Sen black hole in plasma background}

%\keywords{Stringy black hole: gravitational lensing, frame dragging, material medium approach}        % if too long for running head


\author{Saswati Roy}
\email{sr.phy2011@yahoo.com}
\affiliation{Department of Physics, National Institute of Technology, Agartala, Tripura--799046, India}

\author{Shubham Kala}
\email{shubhamkala871@gmail.com}
\affiliation{The Institute of Mathematical Sciences, C.I.T. Campus, Taramani, Chennai--600113, Tamil Nadu, India}

\author{Sayanika Modak}
\email{sayanikamodak1999@gmail.com}
\affiliation{Department of Physics, National Institute of Technology, Agartala, Tripura--799046, India}

\author{Hemwati Nandan}
\email{hnandan@associated.iucaa.in}
\affiliation{Department of Physics, Hemvati Nandan Bahuguna Garhwal Central University, Srinagar, Garhwal, Uttarakhand--246174, India}
\affiliation{Centre for Space Research, North-West University, Potchefstroom 2520, South Africa}

\author{Amare Abebe}
\email{Amare.Abebe@nithecs.ac.za}
\affiliation{Centre for Space Research, North-West University, Potchefstroom 2520, South Africa}
\affiliation{National Institute for Theoretical and Computational Sciences (NITheCS), South Africa}

\begin{abstract}
We investigate the gravitational lensing of massless particles around a Kerr–Sen black hole immersed in  a magnetized, cold, pressureless plasma medium. Both homogeneous and inhomogeneous plasma distributions are considered in this study to mimic realistic astrophysical environments. The light deflection angle is computed, and the effects of the black hole’s rotation and charge on light bending are analyzed in detail. The conditions for the circular photon orbits are also examined in both plasma configurations.
A comparison with the vacuum case (i.e. zero plasma frequency) highlights the role of plasma in modifying the light propagation, and the results obtained provide a deeper insight into plasma effects which improve our understanding of observational signatures of rotating charged black holes.
\end{abstract}

\maketitle

%\tableofcontents

\section{Introduction}
The first ever detection of a black hole shadow has significantly enhanced our understanding of the most enigmatic and compact objects in the universe~\cite{EventHorizonTelescope:2019dse}. From the early theoretical predictions of Einstein’s General Theory of Relativity (GR) to the groundbreaking observational evidence provided by the Event Horizon Telescope Collaboration, the study of black holes has entered a remarkable new era~\cite{Einstein:1916vd,schwarzschild1916gravitationsfeld}. The bending of light in the presence of a strong gravitational field, widely known as gravitational lensing, is one of the most important consequences of GR. It allows astronomers to probe the spacetime geometry around these ultra-compact objects, providing direct observational insights into the behavior of matter and radiation under extreme gravity~\cite{Palatini:1923dza,ohanian1987black,schneider1992gravitational,Bekenstein:1993fs,Kopeikin:2006hy}. The first experimental confirmation of light bending was achieved during Eddington’s 1919 expedition, providing one of the earliest observational proofs of GR \cite{Dyson:1920cwa}. Since then, the study of gravitational lensing has evolved significantly, with contributions from many researchers. The behavior of light near the photon sphere, showing a logarithmic divergence in the deflection angle as first recognized by Darwin, established the basis for strong gravitational lensing studies~\cite{darwin1959gravity}. Refsdal demonstrated that gravitational fields can bend light to generate multiple images of a background source, providing the theoretical basis for lensing investigations~\cite{Refsdal:1964yk}. Later, Iyer developed a perturbative framework that extended the strong-deflection bending angle to include spin-dependent effects for rotating black holes~\cite{Iyer:2006cn}. Furthermore, Virbhadra and Ellis revived interest in black hole lensing, leading to the widely adopted Virbhadra–Ellis lens equation~\cite{virbhadra2000schwarzschild,Virbhadra:2008ws}. A significant advancement was subsequently made by Bozza, who introduced the strong deflection limit formalism, enabling analytical calculations of deflection angles and key lensing observables such as image positions, angular separations, and magnifications for a wide range of spherically symmetric black hole spacetimes~\cite{Bozza:2001xd,Bozza:2002zj}. In particular, Iyer \textit{et al.} also derived spin-dependent corrections for rotating black holes, showing asymmetric deflection for direct and retrograde photon orbits and corresponding shifts in strong-lensing image positions~\cite{Iyer:2009wa}. Additionally, Gibbons and Werner proposed a geometric and topological approach based on the Gauss–Bonnet theorem for studying light deflection in static, spherically symmetric spacetimes~\cite{Gibbons:2008rj}. Subsequently, Ono \textit{et al.}~\cite{Ono:2018ybw} generalized this method to stationary, axisymmetric, and asymptotically flat spacetimes, while Arakida investigated finite-distance deflection effects~\cite{Arakida:2017hrm}. More recently, Huang \textit{et al.} refined the Gauss–Bonnet approach to derive simplified expressions for the deflection angle, and further generalizations have extended weak-field formulas to non-asymptotically flat geometries~\cite{Huang:2023bto}.

In a purely vacuum environment, the bending of light is determined entirely by the gravitational field of the lensing mass and is independent of the photon’s wavelength. In contrast, most astrophysical settings contain ionized plasma, which modifies photon trajectories due to its dispersive nature, making light deflection frequency-dependent~\cite{Synge:1966okc}. In such plasmas, photons no longer strictly follow null geodesics; instead, their motion can be effectively described using a spatially and frequency-dependent refractive index. Ehlers and his collaborators formulated a Hamiltonian description for light rays propagating through a pressureless magnetized plasma in curved spacetime~\cite{Ehlers:1987nm}. Building on this framework, Perlick and his colleagues calculated deflection angles around Schwarzschild and Kerr black holes for plasmas with radial density profiles~\cite{Perlick:2004tq,Perlick:2017fio}. Bisnovatyi-Kogan and Tsupko further provided comprehensive studies of lensing effects in plasma environments~\cite{Bisnovatyi-Kogan:2010flt,tsupko2012gravitational,Tsupko:2013cqa}. Morozova and collaborators extended these analyses to slowly rotating Kerr black holes restricted to the equatorial plane~\cite{morozova2013346}. More recent research, including work by Er, Mao, and Rogers, has investigated the wavelength-dependent bending of light in a variety of realistic astrophysical scenarios~\cite{Er:2013efa,rogers2015frequency}. So far, studies of gravitational lensing in the presence of plasma around black holes have been carried out for a range of alternative theories of gravity, with several recent examples cited here, illustrating the growing interest in understanding lensing signatures beyond standard GR~\cite{Crisnejo:2018uyn,Atamurotov:2021hoq,Atamurotov:2021qds,Javed:2021ymu,Kala:2022uog,Li:2023esz,Atamurotov:2023rye,Kala:2024fvg,Feng:2024jeq,Rahmatov:2025gpk,Hoshimov:2025hyt,Kala:2025fld,Turakhonov:2025ojy,Mustafa:2025cou}.

This work investigates gravitational lensing in a plasma environment around the Kerr-Sen black hole~(KSBH), a solution arising from heterotic string theory~\cite{Gross:1984dd}. The KSBH describes a rotating, charged black hole and represents a dilaton–axion extension of the classical Kerr black hole in GR~\cite{Garfinkle:1990qj,Sen:1994eb}. Its rich structure and theoretical significance have made it a subject of growing interest in recent years. Prior studies have investigated the gravitational deflection of light around Kerr–Sen black holes. In particular, Uniyal \textit{et al.}~\cite{uniyal2018bending} performed a detailed analysis in the equatorial plane and derived exact expressions for the light deflection angle in both weak and strong-field regimes in vacuum. Furthermore, the deflection of light due to KSBH using a material medium was examined in greater detail by Roy \textit{et al.}~\cite{Roy:2025hdw}. Recent investigations have highlighted the rich observational phenomenology of KSBH in different astrophysical environments. Hotspot images driven by magnetic reconnection processes and their observable signatures have been explored in detail~\cite{Wang:2025buh}. In addition, the optical appearance of a KSBH surrounded by a thin accretion disk has been systematically analyzed~\cite{Wang:2025dyw}. In addition to these studies, shadow signatures of KSBH embedded in a plasma medium have been identified using Event Horizon Telescope data, providing important links between theoretical models and observational constraints~\cite{KumarSahoo:2025igt}. Building on these investigations, the present study focuses on the effects of a plasma environment on gravitational lensing by KSBH.

The structure of this manuscript is organized as follows: In Section~II, the KSBH spacetime geometry is presented in detail, along with a discussion of frame-dragging effects.
Section~III is devoted to the equations of motion for light propagation in a plasma medium and the corresponding deflection angle in different plasma environments. In Section~IV, the conditions for circular photon orbits are analyzed for the considered plasma configurations. Finally, Section~V summarizes the main results and presents the conclusions.

%%%%%%%%%%_______________________________section______________________________%%%%%%%%%%%%%%%%%%%
\section{Kerr Sen Black Hole}
A four-dimensional solution of the Kerr black hole (KBH) arising from heterotic string theory, which describes a rotating, electrically charged massive entity, is known as the KSBH solution. In this view, the appropriate effective action\cite{schneider1992gravitational} is as follows:
\begin{equation} \label{Eq:1}
    S =\int_{}^{}d^4x \sqrt{|\Tilde{g}|}e^{-\Tilde{\Phi}}(R-\frac{1}{12}H^2 +\Tilde{g}^{\mu\nu}\partial_\mu\Tilde{\Phi}\partial_\nu\Tilde{\Phi}- \frac{1}{8}F^2),
\end{equation}
where $g$ is the determinant of the metric tensor $g_{\mu\nu}$, $F^2$ is the square of the field-strength tensor $F_{\mu\nu}=\partial_\mu A_\nu -\partial_\nu A_\mu$, $\Phi$ is the dilaton; $H^2$
is the square of a rank-3 tensor field:
\begin{equation} \label{Eq:2}
\begin{split}
    H_{\kappa\mu\nu}=&\partial_\kappa B_{\mu\nu} + \partial_\nu B_{\kappa\mu} + \partial_\mu B_{\nu\kappa}\\
    & -\frac{1}{4}(A_\kappa F_{\mu\nu} + A_\mu F_{\nu\kappa} + A_\nu F_{\kappa \mu}),
\end{split}
\end{equation}
where $B_{\mu\nu}$ is a rank-2 anti-symmetric tensor field. It is evident that the Einstein-Hilbert action results when all non-gravitational fields in the action \eqref{Eq:1} vanish. The Kerr-Sen metric is a solution in the theory given by \eqref{Eq:1} when all non-gravitational fields are absent since it satisfies the vacuum Einstein equations. Thus, the KSBH solution is produced \cite{siahaan2016destroying} by implementing a series of transformations that link the solution in \eqref{Eq:1} to the KBH metric.
 
In a dilaton field, the precise solution to Einstein's Field Equations of GR for a stationary, axially symmetric gravitational field of a rotating body can be provided by the KSBH metric. In the Boyer-Lindquist coordinates ($ct, r, \theta, \phi$), the KSBH line element can be represented as follows \cite{siahaan2016destroying}:
\begin{equation} \label{Eq:3}
\begin{split}
    ds^2 = &-\frac{\Delta-\alpha^2 \sin^2{\theta}}{\Sigma^2}c^2dt^2 + \frac{\Sigma^2}{\Delta}dr^2 + \Sigma^2d\theta^2 \\
    &+\left(\Delta +\frac{r_gr(r(r+2b)+\alpha^2)} {\Sigma^2}\right) \sin^2{\theta} d\phi^2 \\& - \frac{2r_gr\alpha\sin^2{\theta}}{\Sigma^2}cd\phi dt, 
\end{split}
\end{equation}
where the metric functions are described as follows:
\\
\begin{equation} \label{Eq:4}
\begin{split}
    &\Sigma^2 =r(r+2b)+\alpha^2\cos^2{\theta}, \\
    &\Delta =r(r+2b)-r_gr+\alpha^2, \\
    &b =\frac{Q^2}{2m}=\frac{Q^2}{r_g},\\
    &\alpha =\frac{J}{Mc}, \\
    &r_g =\frac{2GM}{c^2}. 
\end{split}
\end{equation}
The symbol $Q$ (in the dimension of length) represents the electric charge of BH, BH's mass and rotational parameter (in the dimension of length) are expressed by $M$ and  $\alpha$ respectively. The constant $r_g=2m$, with $m\equiv GM/c^2$, is known as the Schwarzschild radius. The coefficients of the line element (\ref{Eq:3}) exhibit independence from $\phi$, indicating its axial symmetry. In the theory represented by the action (\ref{Eq:1}), the solutions for non-gravitational basic fields are as follows \cite{siahaan2016destroying}:

\begin{eqnarray} \label{Eq:5}
   && \Tilde{\Phi}=-\frac{1}{2}\ln{\frac{\Sigma^2}{r^2+\alpha^2\cos^2{\theta}}},\\
&& \label{Eq:6}
    A_t=\frac{-rQ}{\Sigma^2},\\
&& \label{Eq:7}
    A_\phi=\frac{rQ\alpha \sin^2{\theta}}{\Sigma^2}\;,\\
&&\label{Eq:8}
    B_{tQ}=\frac{br\alpha \sin^2{\theta}}{\Sigma^2}\;.
\end{eqnarray}
\\
Setting the parameter $b = 0$ yields the result that all non-gravitational fields (\ref{Eq:5} - \ref{Eq:8}) vanish, and the line element (\ref{Eq:3}) reduces to the Kerr metric. Furthermore, if the rotation parameter disappears (i.e $\alpha = 0$), the line element (\ref{Eq:3}) is reduced to the Schwarzschild black hole (SBH) metric, which represents the electrically neutral, non-rotating, or static gravitating mass. But if only the rotational parameter disappears as $\alpha = 0$, the line element (\ref{Eq:3}) is reduced to the Gibbons–Maeda–Garfinkle–Horowitz–
Strominger black hole (GMGHSBH), which represents the charged string black hole.

The spherical event horizon of the Kerr-Sen line element (\ref{Eq:3}) is determined by $g_{tt}=0$ which implies $\Delta(r)=0$, whose outer and inner horizons are located at
\begin{equation} \label{hor_out_in}
   r_\pm = m-b\pm\sqrt{(m-b)^2-\alpha^2}.\\
\end{equation}
 Eq. (\ref{hor_out_in}) yields $(m-b)\geq \alpha$ or $(m- \frac{Q^2}{r_g})\geq \alpha$ unless the horizon disappears. The ranges for the rotation parameter ($\alpha$) and charge ($Q$) for KSBH are $0\leq \alpha \leq m$ and $0\leq Q \leq \sqrt{2} m$. The horizon structure of KSBH is greatly illustrated and compared with other well known BHs in \cite{Roy:2025hdw}.

The linearized form of the KSBH (\ref{Eq:3}) can be written in terms of spherical polar coordinates ($ct, r, \theta, \phi$) as  \cite{Roy:2025hdw}
\begin{equation} 
\begin{split}
    ds^2 = &-\left(\frac{\Delta-\alpha^2 \sin^2{\theta}}{\Sigma^2} + \frac{2r_gr\alpha \sin^2{\theta}}{\Sigma^2} \frac{d\phi}{cdt}\right)c^2dt^2  + \frac{\Sigma^2}{\Delta}dr^2 \\ 
    &+ \Sigma^2d\theta^2  + \left(\Delta + \frac{r_gr(r(r+2b)+\alpha^2)}{\Sigma^2}\right) \sin^2{\theta} d\phi^2\\
    = & -\left(1 -\frac{r_g r }{\Sigma^2} + \frac{2r_gr\alpha \sin^2{\theta}}{\Sigma^2} \frac{d\phi}{cdt}\right)c^2dt^2 + \frac{\Sigma^2}{\Delta}dr^2 \\
     & + \Sigma^2d\theta^2  
    + \left(\Delta + \frac{r_gr(r(r+2b)+\alpha^2)}{\Sigma^2}\right) \sin^2{\theta} d\phi^2.
\end{split}
\end{equation}
Under far field approximation we can assume $\frac{\alpha^2}{r^2}<<1$ hence we get the line element as  \cite{Roy:2025hdw}
\begin{equation} 
\begin{split}
   ds^2 = & -\left(1 -\frac{r_g  }{(r+2b)} + \frac{2r_g\alpha \sin^2{\theta}}{(r+2b)} \frac{d\phi}{cdt}\right)c^2dt^2 \\ 
    & + \frac{(r+2b)}{(r+2b-r_g )}dr^2 + r(r+2b) d\Omega^2\;,
\end{split}
\end{equation}
where $d\Omega^2=d\theta^{2}+\sin^{2}\theta d\phi^{2}$ and $\frac{d\phi}{cdt}$ is the frame-dragging parameter, which arises as a result of the rotation of the gravitating mass.

\subsection{\textbf{Frame-dragging}}  
The relativistic action function $S$, for a particle with time $t$ and angle $\phi$ as cyclic variables, in the gravitational field of a rotating spherical mass ~\cite{landau1951d}, is as follows:
\begin{equation} \label{Eq_action}
    S=-E_0t+L\phi+S_r(r)+S_\theta(\theta)\;.
\end{equation}
In the above equation $E_0$ denotes the conserved energy, while $L$ represents the part of the angular momentum along the field's symmetry axis and $S_r$, $S_\theta$ represent the parts of the action associated with the radius and the polar angle respectively.\\ 
The four momentum of the particle can be written as
\begin{equation} \label{Eq_four_momentum}
     p^i=mc\frac{dx^i}{ds}=g^{ik}p_k=-g^{ik}\frac{\partial S}{\partial x^k}.
\end{equation}
Here, i and k have values that go from 0 to 3 which stand for
the coordinates $ct$, $r$, $\theta$ , $\phi$ respectively \cite{landau1951d}. 

So, using Eq. (\ref{Eq_four_momentum}), the general expression of the frame-dragging for KSBH ($\frac{d\phi}{cdt}$) is obtained as \cite{Roy:2025hdw},
\begin{equation} \label{Eq:dphi_cdt}
\begin{split}
\frac{d\phi}{cdt}=&
        \frac{r_gr\alpha sin^2{\theta} \frac{E_0}{c}+(\Sigma^2-r_gr)L}{\left[\left\{\Sigma^2(r(r+2b)+\alpha^2)+r_gr\alpha^2 sin^2{\theta}\right\}\frac{E_0}{c}-r_gr\alpha L\right]}\\& \times \frac{1}{sin^2{\theta}}.
   \end{split}
\end{equation}

We express the angular momentum $L$ as $L=p\beta$,
where $\beta$ is the impact parameter
and for light-like particles (photons), the momentum ($p$) and the conserved energy ($E_0$) are related as $E_0=pc $
which yields $\beta=\frac{Lc}{E_0}$.  
Under the far-field approximation, $\frac{\alpha^2}{r^2}<<1$ so that $\Sigma^2=r(r+2b)$ and the above equation becomes \cite{Roy:2025hdw}
\begin{equation} \label{Eq_dphidt}
\begin{split}
    \frac{d\phi}{c dt}&=\frac{r_g r\alpha \sin^2{\theta} +\{r(r+2b)-r_g r\}\beta}{r^2(r+2b)^2-r_g r\alpha\beta} \times \frac{1}{sin^2{\theta}}\\
    &=\frac{r_g\alpha \sin^2{\theta} +\{(r+2b)-r_g\}\beta}{r(r+2b)^2-r_g\alpha\beta} \times \frac{1}{sin^2{\theta}}.\\
\end{split}
\end{equation}
Thus, the Kerr-Sen metric in spherical symmetric form can be expressed as \cite{Roy:2025hdw} 

\begin{widetext}
\begin{equation}
ds^2 = -\left(1 - \frac{r_g}{(r+2b)} + \frac{2 r_g \alpha \sin^2{\theta}}{(r+2b)} 
\times \frac{r_g \alpha \sin^2{\theta} + \{(r+2b)-r_g\}\beta}{r(r+2b)^2 - r_g \alpha \beta} 
\times \frac{1}{\sin^2{\theta}} \right)c^2 dt^2 
+ \frac{(r+2b)}{(r+2b-r_g)} dr^2 
+ r(r+2b)\, d\Omega^2 .
\end{equation}
%\end{strip}
%\begin{equation} 
%\begin{split}
%    ds^2 = & -\left(1 -\frac{r_g  }{(r+2b)} + \frac{2r_g\alpha \sin^2{\theta}}{(r+2b)}  \times \frac{r_g\alpha \sin^2{\theta} +\{(r+2b)-r_g\}\beta}{r(r+2b)^2-r_g\alpha\beta} \times \frac{1}{sin^2{\theta}} \right)c^2dt^2 \\ 
%    & + \frac{(r+2b)}{(r+2b-r_g )}dr^2 + r(r+2b) d\Omega^2.
%\end{split}
%\end{equation}
We restrict the analysis to the equatorial plane ($\theta=\pi/2$), so that the Kerr-Sen metric becomes \cite{Roy:2025hdw}
\begin{equation}
ds^2 = -\left(1 - \frac{r_g}{(r+2b)} + \frac{2 r_g \alpha \big[r_g \alpha + \{(r+2b)-r_g\}\beta\big]}{(r+2b)\{r(r+2b)^2 - r_g \alpha \beta\}} \right)c^2 dt^2
+ \frac{(r+2b)}{(r+2b-r_g)} dr^2
+ r(r+2b)\, d\Omega^2 .
\end{equation}
%\begin{equation} 
%\begin{split}
%    ds^2 = & -\left(1 -\frac{r_g  }{(r+2b)} + \frac{2r_g\alpha }{(r+2b)}  \times \frac{r_g\alpha +\{(r+2b)-r_g\}\beta}{r(r+2b)^2-r_g\alpha\beta} \right)c^2dt^2 \\ 
%    & + \frac{(r+2b)}{(r+2b-r_g )}dr^2 + r(r+2b) d\Omega^2,\\
%    = & -\left(1 -\frac{r_g  }{(r+2b)} + \frac{2r_g\alpha [r_g\alpha +\{(r+2b)-r_g\}\beta]}{(r+2b)\{r(r+2b)^2-r_g\alpha\beta\}} \right)c^2dt^2 \\ 
%    & + \frac{(r+2b)}{(r+2b-r_g )}dr^2 + r(r+2b) d\Omega^2.
%\end{split}
%\end{equation}
\end{widetext}
%%%%%%%%%%_______________________________section______________________________%%%%%%%%%%%%%%%%%%%
\section{Equation of motion of Light in Plasma Medium}
The spherically symmetric and static metric is given as
\begin{equation} \label{spherical metric}
	g_{ik}dx^{i}dx^{k} = -A(r)dt^{2} + B(r)dr^{2} +D(r) d\Omega^2,
\end{equation} 
where $A(r)$,  $B(r)$, and $D(r)$ are positive, outside the event horizon.
We investigate the propagation of photons in a non-magnetized, cold (pressureless) electron plasma surrounding a KSBH. Here, cold or pressureless means that there is no significant thermal agitation of the electrons, so the plasma pressure is negligible compared to its rest-mass energy density. This assumption removes the need to include temperature-dependent corrections, allowing us to isolate and study purely the dispersive effects of the plasma on photon motion without introducing cumbersome kinetic terms.

Plasma is characterized by a spatially dependent electron number density \(N(r)\), which defines plasma frequency as~\cite{Synge:1966okc}
\begin{equation}
\omega_p^2(r) = \frac{4\pi e^2}{m_e} \, N(r),
\end{equation}
where \(e\) is the elementary charge and \(m_e\) is the electron rest mass. The quantity \(\omega_p(r)\) sets the natural oscillation frequency of the plasma electrons and serves as a cutoff frequency for electromagnetic wave propagation. Photons with local frequencies below \(\omega_p(r)\) are unable to propagate and instead undergo evanescent decay.
In the presence of gravity, photon motion in plasma deviates from null geodesics of the background spacetime. For a photon with conserved energy \(\omega_0 = -p_t\), measured at spatial infinity, the local frequency is redshifted according to~\cite{Perlick:2004tq},
\begin{equation}
\omega(r) = \frac{\omega_0}{\sqrt{A(r)}},
\end{equation}
where \(A(r)\) is the lapse function of the static, spherically symmetric metric
given in Eq. \ref{spherical metric}.
The corresponding refractive index in plasma is
\begin{equation}
n^2(r,\omega) = 1 - \frac{\omega_p^2(r)}{\omega^2(r)},
\end{equation}
which reduces to unity in vacuum. The propagation condition
\begin{equation}
\omega(r) > \omega_p(r)
\end{equation}
ensures that the photon frequency measured locally exceeds the local plasma frequency. It should be emphasized that the marginal case \(\omega(r) \gtrsim \omega_p(r)\) is of primary interest here, since this is where dispersive effects are significant. The limit \(\omega(r) \gg \omega_p(r)\) corresponds to an effectively vacuum-like situation in which plasma effects become negligible.

Spherical symmetry permits restricting the analysis to the equatorial plane \((\theta = \pi/2)\), which is adopted throughout this study. In the geometric-optics approximation, the photon motion in plasma is described by the Hamiltonian~\cite{tsupko2012gravitational}
\begin{equation}
H = \frac12 \left[ g^{\mu\nu} p_\mu p_\nu + \omega_p^2(r) \right] = 0.
\end{equation}
Substituting the metric components gives
\begin{equation}
-\frac{\omega_0^2}{A(r)} + \frac{p_r^2}{B(r)} + \frac{L^2}{D(r)} + \omega_p^2(r) = 0,
\end{equation}
where \(L = p_\varphi\) is the conserved angular momentum of the photon.

The radial trajectory follows from the relation
\begin{equation}
\frac{dr}{d\varphi} = \frac{D(r)}{B(r)} \frac{p_r}{L}.
\end{equation}
Eliminating \(p_r\) using the Hamiltonian constraint yields the orbit equation as
\begin{equation}
\frac{dr}{d\varphi} = \pm \frac{\sqrt{D(r)}}{\sqrt{B(r)}} 
\sqrt{\frac{\omega_0^2}{L^2} \mathcal{H}^2(r) - 1},
\end{equation}
where the effective function
\begin{equation}
\mathcal{H}^2(r) = \frac{D(r)}{A(r)} \left[1 - A(r)\left(\frac{\omega_p^2(r)}{\omega_0^2}\right)\right]
\end{equation}
encapsulates the combined influence of spacetime curvature and plasma dispersion. In the absence of plasma \((\omega_p=0)\), this reduces to the purely geometric case of vacuum photon motion in GR.

\subsection{\textbf{Light Deflection in Homogeneous Plasma medium}}

The deflection angle can be expressed as a function of only $R$ and $\omega_{0}$ (for a given plasma distribution) as follows:
\begin{equation}
	\delta_{h} + \pi = 2 \int_{R}^{\infty} \frac{\sqrt{B(r)}}{\sqrt{D(r)}} \left[\frac{\mathcal{H}^{2}_{h}(r)}{\mathcal{H}^{2}_{h}(R)} -1 \right]^{-1/2} dr\;.
\end{equation}
In the present case, 
\begin{equation}
\begin{split}
    A(r) = & \left(1 -\frac{r_g  }{(r+2b)} + \frac{2r_g\alpha [r_g\alpha +\{(r+2b)-r_g\}\beta]}{(r+2b)\{r(r+2b)^2-r_g\alpha\beta\}} \right)c^2,\\
    B(r) = & \frac{(r+2b)}{(r+2b-r_g )},\\
    D(r) = & r(r+2b).
\end{split}
\end{equation}
\begin{widetext}
The expression for $\mathcal{H}_{h}(r)$ for the Kerr-Sen spacetime is

 \begin{equation} \label{HrhomoPlas}
   \mathcal{H}^{2}_{h}(r)= r(r+2b) \left[\frac{(r+2b)\{r(r+2b)^2-r_g\alpha\beta\}}{(r+2b-r_g)[r(r+2b)^2+r_g\alpha\beta]+2 r_g^2\alpha^2}- \frac{\omega_p^2(r)}{\omega_0^2}\right]\;.
\end{equation}   
\end{widetext}

 %%%%%%%%%%%%%%%%%%%%%%%%%%%%%%%%%%%%%%%%%%%%%%%%%%
\begin{figure}[htbp]
	\begin{center}
		{\includegraphics[width=0.45\textwidth]{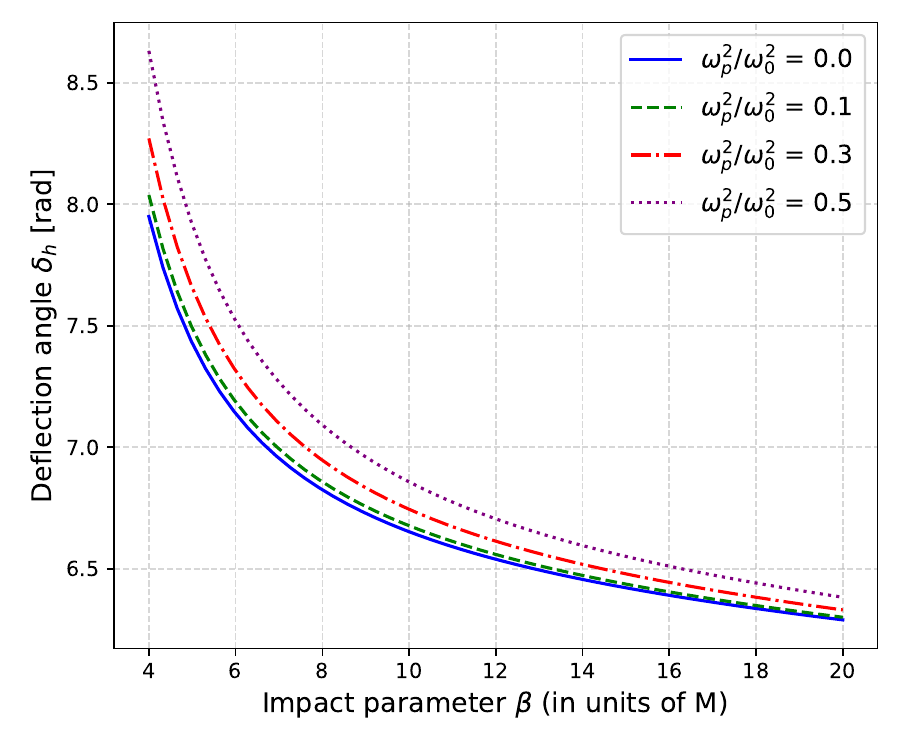}}
      {\includegraphics[width=0.45\textwidth]{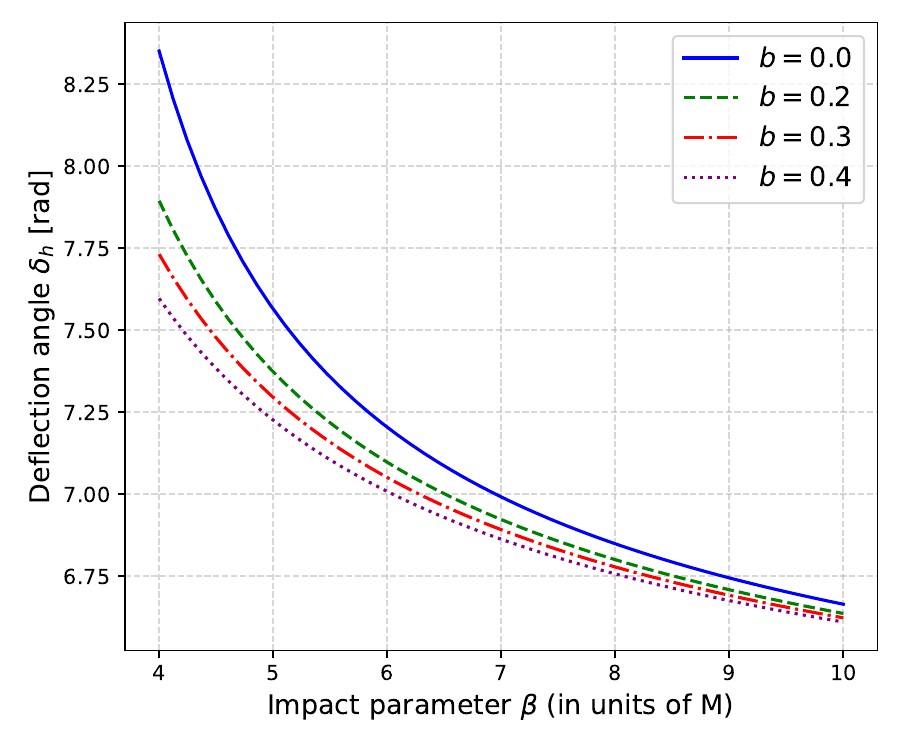}}
       {\includegraphics[width=0.45\textwidth]{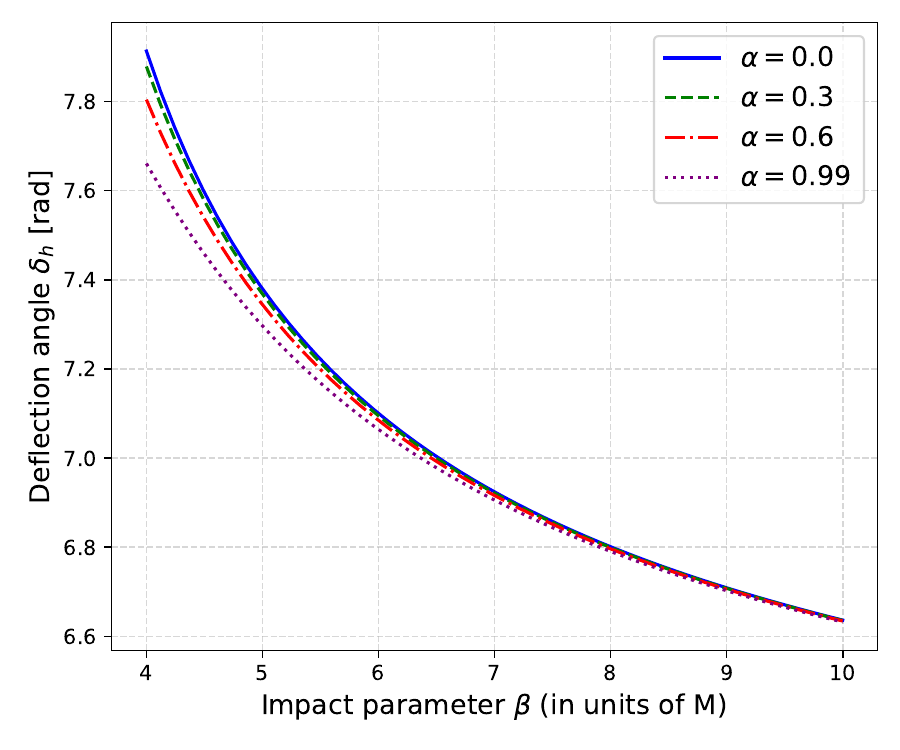}}
	\end{center}
	\caption{Behavior of the deflection angle as a function of the impact parameter in a homogeneous plasma medium. The upper panel shows the effect of changing the homogeneous plasma parameter $\omega_p^2/\omega_0^2$ for fixed values of $\alpha = 0.5$ and $b = 0.5$. The middle panel illustrates the dependence on different values of $b$ while keeping $\alpha = 0.2$ and $\omega_p^2/\omega_0^2 = 0.5$ constant. The lower panel presents the variation with different values of $\alpha$ for fixed $b = 0.2$ and $\omega_p^2/\omega_0^2 = 0.5$.}\label{fig:HPDA1}
\end{figure}
%%%%%%%%%%%%%%%%%%%%%%%%%%%%%%%%%%%%%%%%%%%%%%%%%%
 %%%%%%%%%%%%%%%%%%%%%%%%%%%%%%%%%%%%%%%%%%%%%%%%%%
\begin{figure}[htbp]
	\begin{center}
		{\includegraphics[width=0.48\textwidth]{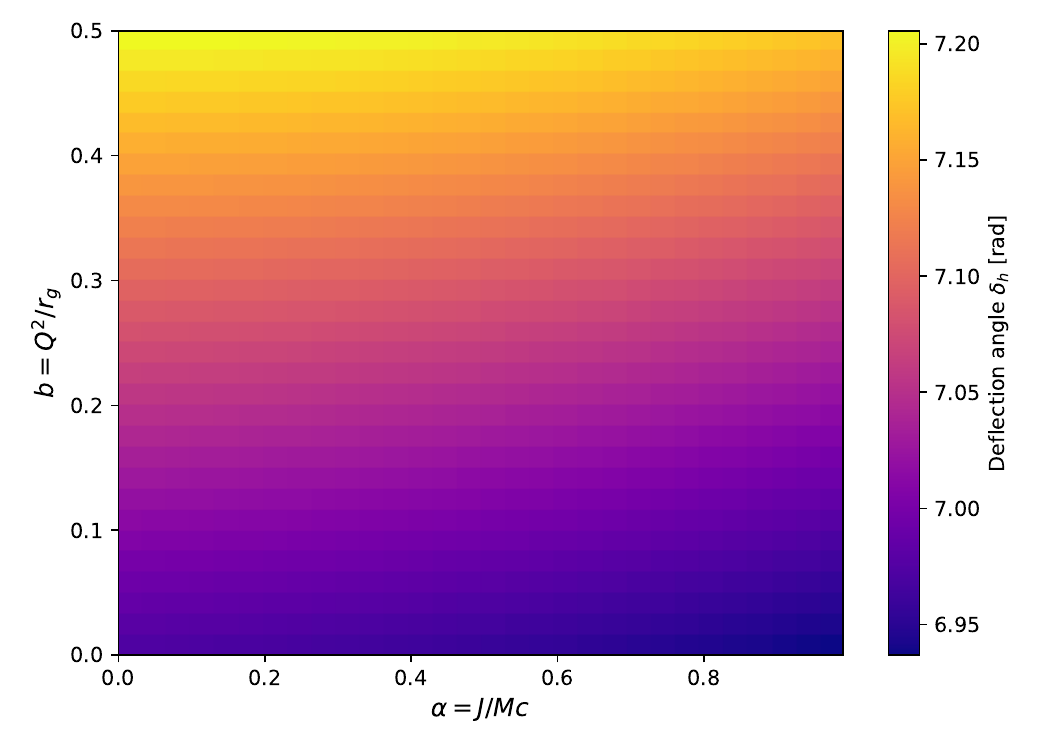}}
	\end{center}
	\caption{Density plot showing the variation of the deflection angle as a function of the spin parameter $\alpha$ and the charge parameter $b$ in a homogeneous plasma medium. The plasma frequency ratio is fixed at $\omega_p^2/\omega_0^2 = 0.5$, with $b = 6$ and $M = 1$. The color scale represents the magnitude of the deflection angle, illustrating how it changes across the $(\alpha, b)$ parameter space.}
\label{fig:HPDA2}
\end{figure}
Fig.~\ref{fig:HPDA1} illustrates the variation of the deflection angle $\delta_{h}$ as a function of the impact parameter $\beta$ in a homogeneous plasma medium for different values of the black hole parameters. In the upper panel, the effect of varying the homogeneous plasma parameter $\omega_{p}^{2}/\omega_{0}^{2}$ is shown for fixed values of $\alpha = 0.5$ and $b = 0.2$. It is seen that the deflection angle increases with larger plasma frequency ratios, which physically corresponds to the fact that a denser plasma medium slows down light propagation and enhances the bending of photon trajectories. The middle panel shows the dependence of the deflection angle on the charge parameter $b$, keeping $\alpha = 0.2$ and $\omega_{p}^{2}/\omega_{0}^{2} = 0.5$ constant. In this case, increasing the charge reduces the deflection angle, indicating that the repulsive contribution of the black hole’s charge counteracts the gravitational focusing of photons. The lower panel presents the variation with the rotation parameter $\alpha$ for fixed $b = 0.2$ and $\omega_{p}^{2}/\omega_{0}^{2} = 0.5$. The results show that higher rotation diminishes the deflection angle, which can be attributed to the frame-dragging effects of the Kerr--Sen geometry altering photon trajectories and weakening the net bending. To complement these results, Fig.~\ref{fig:HPDA2} shows the simultaneous influence of both the charge and the rotation parameter on the deflection angle in the form of a density plot. This representation makes it clear that the combined effect of increasing charge and rotation leads to a monotonic decrease in the bending of light, emphasizing the competing roles of plasma dispersion, black hole charge, and spin in determining gravitational lensing signatures. Together, these figures highlight the rich interplay of physical effects governing photon propagation in the Kerr--Sen spacetime and provide valuable insights into possible observational imprints of charged, rotating black holes embedded in a plasma environment.
%%%%%%%%%%%%%%%%%%%%%%%%%%%%%%%%%%%%%%%%%%%%%%%%%%
\subsection{\textbf{Light Deflection in Non-Homogeneous Plasma Medium}}
In the following, we extend our analysis to the case of a non-homogeneous plasma medium, 
where the plasma frequency varies with the radial coordinate according to a power-law distribution~\cite{rogers2015frequency}, 
\begin{equation}
    \omega^{2}_{p}(r) = \frac{\eta_{0}}{r^{\nu}},
\end{equation}
where $\eta_{0}$ denotes the plasma strength and $\nu$ $(\geq 0)$  is the radial decay index.
This profile captures more realistic astrophysical scenarios in which the plasma density decreases
with increasing distance from the black hole. Incorporating this dependence modifies the effective
optical geometry probed by photons. In particular, the trajectory function $\mathcal{H}(r)$ is modified to
\begin{equation} \label{HrnonhomoPlas}
    \mathcal{H}^{2}_{nh}(r) = \frac{D(r)}{A(r)} \left[ 1 - \frac{A(r)\,\eta_{0}}{\omega_{0}^{2}\, r^{\nu}} \right],
\end{equation}
which makes the bending angle explicitly sensitive to both the amplitude ($\eta_{0}$) 
and the radial fall-off rate ($\nu$) of the plasma distribution. 

The corresponding deflection angle in this medium can still be expressed in integral form,
\begin{equation}
    \delta_{nh} + \pi = 2 \int_{R}^{\infty} \frac{\sqrt{B(r)}}{\sqrt{D(r)}}
    \left[ \frac{\mathcal{H}^{2}_{nh}(r)}{\mathcal{H}^{2}_{nh}(R)} - 1 \right]^{-\tfrac{1}{2}} \, dr.
\end{equation}
As in the homogeneous case, we have evaluated this expression numerically. 
This framework provides a systematic approach to analyze the influence of different plasma 
density profiles on the deflection of light near compact objects.
 %%%%%%%%%%%%%%%%%%%%%%%%%%%%%%%%%%%%%%%%%%%%%%%%%%
\begin{figure}[htbp]
	\begin{center}
		{\includegraphics[width=0.48\textwidth]{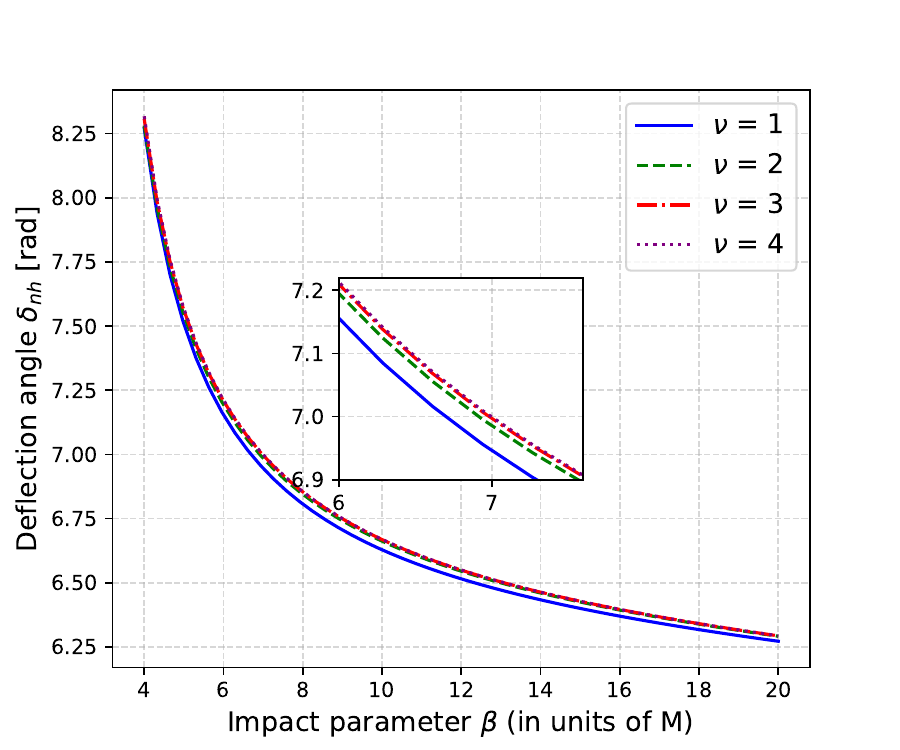}}
      {\includegraphics[width=0.45\textwidth]{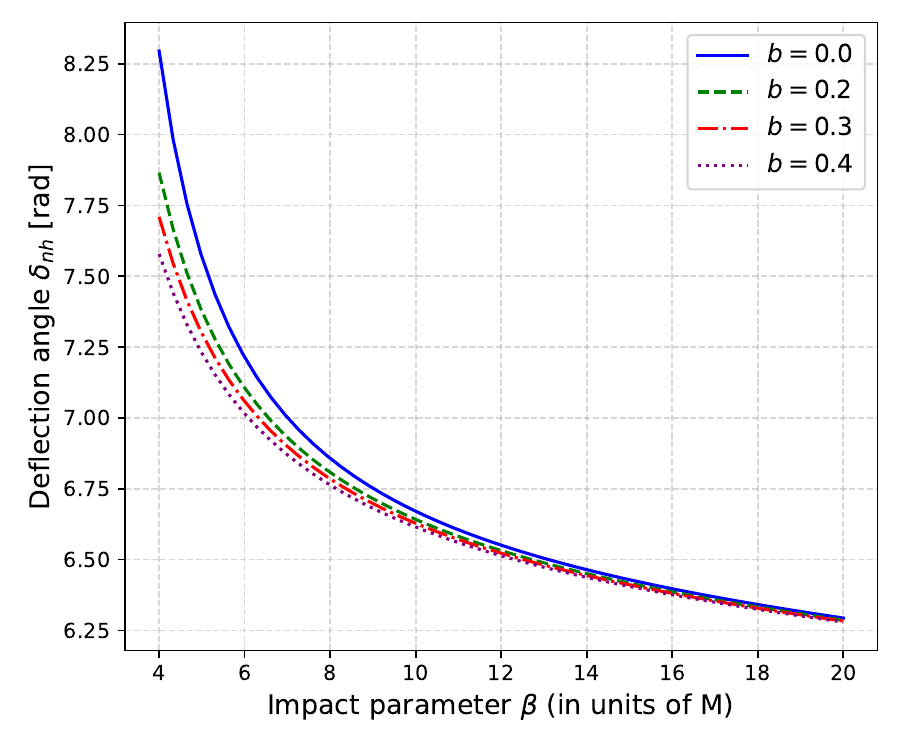}}
       {\includegraphics[width=0.45\textwidth]{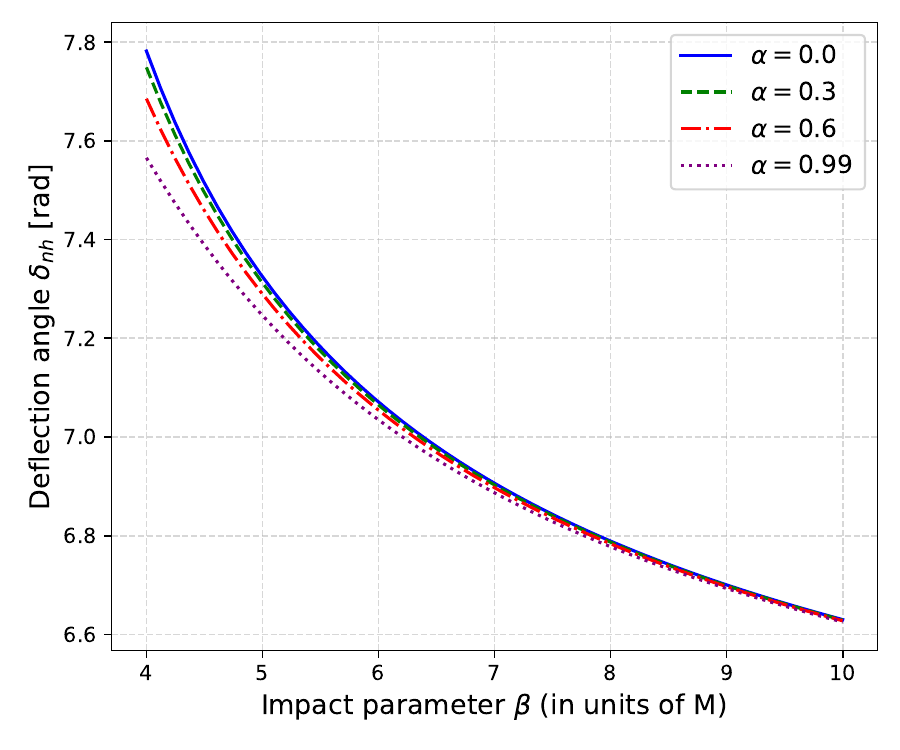}}
	\end{center}
	\caption{Behavior of the deflection angle as a function of the impact parameter in a non-homogeneous plasma medium. The upper panel shows the effect of changing the non-homogeneous plasma index $\nu$ for fixed values of $\eta_{0}=0.5$, $\alpha = 0.5$ and $b = 0.2$. The middle panel illustrates the dependence on different values of $b$ while keeping $\alpha = 0.2$, $\eta_{0}=0.5$ and $\nu = 2$ constant. The lower panel presents the variation with different values of $\alpha$ for fixed $b = 0.2$, $\eta_{0}=0.5$ and $\nu = 2$.}\label{fig:NonHPDA1}
\end{figure}
%%%%%%%%%%%%%%%%%%%%%%%%%%%%%%%%%%%%%%%%%%%%%%%%%%
 %%%%%%%%%%%%%%%%%%%%%%%%%%%%%%%%%%%%%%%%%%%%%%%%%%
\begin{figure}[htbp]
	\begin{center}
		{\includegraphics[width=0.48\textwidth]{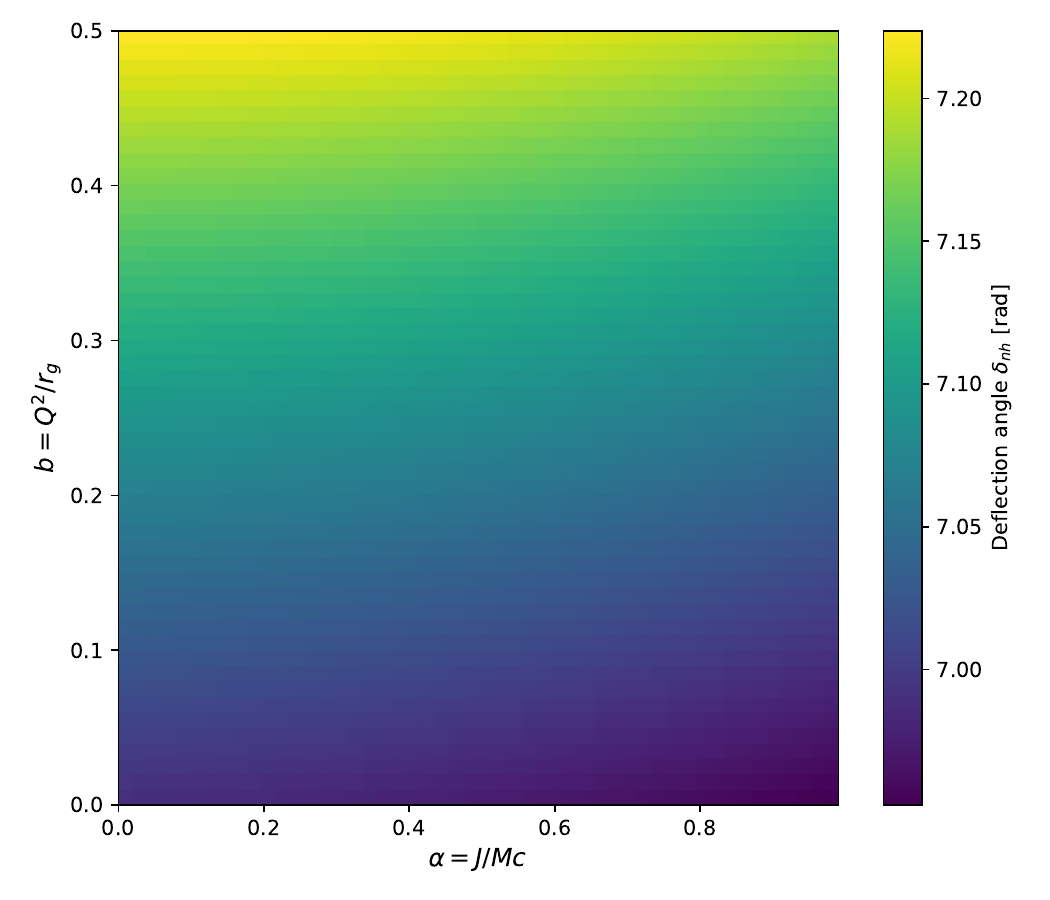}}
	\end{center}
	\caption{Density plot showing the variation of the deflection angle as a function of the spin parameter $\alpha$ and the charge parameter $b$ in a homogeneous plasma medium. The non-homogeneous plasma frequency index is fixed at $\eta_{0}=0.5$, $\nu=2$ with $b = 6$ and $M = 1$. The color scale represents the magnitude of the deflection angle, illustrating how it changes across the $(\alpha, b)$ parameter space.}
\label{fig:NonHPDA2}

\end{figure}
%%%%%%%%%%%%%%%%%%%%%%%%%%%%%%%%%%%%%%%%%%%%%%%%%%
Fig.~\ref{fig:NonHPDA1} shows the variation of the deflection angle $\delta_{nh}$ as a function of the impact parameter $\beta$ in a non-homogeneous plasma medium around a charged, rotating black hole. The upper panel illustrates the impact of the plasma index $\nu$, showing that larger values of $\nu$ lead to a notable increase in the bending of light beyond the critical value of $\beta$. This phenomenon can be attributed to the fact that an increase in $\nu$ steepens the plasma density gradient near the black hole, thereby amplifying the refractive index variation in that region. Moreover, the nearly indistinguishable deflection angles observed for $\nu = 3$ and $\nu = 4$ indicate a saturation effect, suggesting that beyond a specific threshold, further changes in the plasma profile yield no variation in bending. This observation implies that within this parameter range, light propagation is primarily governed by the combined influence of gravitational and plasma effects, with higher-order alterations in the plasma density profile exerting minimal influence on the deflection angle. The middle panel illustrates the effect of the black hole charge $b$, which reduces the deflection angle for fixed $\nu$ and $\alpha$, indicating that the electric charge weakens the gravitational lensing. The lower panel shows the impact of the rotation parameter $\alpha$, where a higher rotation similarly decreases light bending. These results underline the competing interplay between plasma and black hole properties: while the non-homogeneous plasma increases light deflection by modifying the refractive index along the photon path, both rotation and charge tend to reduce it. Fig.~\ref{fig:NonHPDA2} shows the simultaneous influence of both the charge and the rotation parameter on the deflection angle in the form of a density plot for the non-homogeneous plasma case. This interplay produces a complex lensing signature that is markedly different from the homogeneous plasma or vacuum cases, emphasizing the importance of plasma structure in realistic astrophysical scenarios.
%%%%%%%%%%_______________________________section______________________________%%%%%%%%%%%%%%%%%%%
%%%%%%%%%%%%%%%%%%%%%%%%%%%%%%%%%%%%%%%%%%%%%%%%%%%%%%%%%%%%%%%%%%%%%%%%%%%%%%%%
\section{Circular light orbits in plasma environment}
The photon sphere plays a central role in determining the shadow cast by a black hole, as it represents the region where massless particles can move along circular trajectories. First emphasized by Atkinson, the photon sphere marks the boundary between light rays that are scattered back to infinity and those that are inevitably captured. In asymptotically flat geometries, where the plasma frequency satisfies $\omega_{p}(r) \to 0$ as $r \to \infty$, the outermost photon sphere is typically unstable against radial perturbations. As a result, photons that approach it may spiral around the black hole for some time but cannot remain confined indefinitely. For trajectories with a turning point $R > r_{ps}$, the photon eventually escapes to infinity, while in the limiting case $R = r_{ps}$, the photon asymptotically approaches the unstable circular orbit of radius $r_{ps}$. Although certain spacetimes may admit multiple photon spheres, the physically relevant one for shadow formation is the outermost unstable orbit.
In the framework of Perlick et al.~\cite{Perlick:2017fio}, the photon sphere radius $r_{ps}$ is characterized by the stationary behavior of the effective function $\mathcal{H}(r)$, expressed as
\begin{equation}
\frac{d}{dr} \mathcal{H}^{2}(r) = 0.
\end{equation}
The photon sphere radius is obtained by solving the above equation for $r = r_{ps}$. This condition can be evaluated separately for the homogeneous and radially varying plasma distributions discussed above, yielding the corresponding modified photon sphere radius in each case.
\subsection{\textbf{Circular light orbits in homogeneous plasma medium}}
\noindent The effective metric function for a homogeneous plasma, as obtained in Eq.~\ref{HrhomoPlas}, provides the basis for determining the photon sphere. Differentiating $\mathcal{H}^{2}_h(r)$ with respect to the radial coordinate $r$, we find
\begin{multline}
\frac{d}{dr}\mathcal{H}_{h}^{2}(r) = \frac{D'(r)}{A(r)} \Bigl[1 - A(r) \frac{\omega^2_p(r)}{\omega_0^2} \Bigr] \\
- \frac{D(r) A'(r)}{A^2(r)} \Bigl[1 - A(r) \frac{\omega^2_p(r)}{\omega_0^2} \Bigr] \\
- \frac{D(r)}{A(r)} \Bigl[ A'(r) \frac{\omega^2_p(r)}{\omega_0^2} + A(r) \frac{2 \omega_p(r) \omega_p'(r)}{\omega_0^2} \Bigr].
\end{multline}
\noindent Imposing the condition $\frac{d}{dr}\mathcal{H}^{2}_h(r)=0$ at $r=r_{ps}$ determines the location of the photon sphere in the presence of plasma. Explicitly, this condition becomes
\begin{multline}
\frac{2 r_{ps}}{A(r_{ps})} \biggl[ 1 - A(r_{p}) \frac{\omega_p^{2}(r_{p})}{\omega_{0}^{2}} \biggr]
- \frac{r_{ps}^{2} A'(r_{ps})}{A^{2}(r_{ps})} \biggl[ 1 - A(r_{ps}) \frac{\omega_p^{2}(r_{ps})}{\omega_{0}^{2}} \biggr] \\
- \frac{r_{ps}^{2}}{A(r_{ps})} \biggl[ A'(r_{ps}) \frac{\omega_p^{2}(r_{ps})}{\omega_{0}^{2}} 
+ A(r_{ps}) \frac{2 \omega_p(r_{ps}) \omega_p'(r_{ps})}{\omega_{0}^{2}} \biggr] = 0.
\end{multline}
The above equation does not admit a simple closed-form solution without any approximation. Consequently, we solve it numerically in order to explore how the homogeneous plasma distribution modifies the photon sphere radius.
%%%%%%%%%%%%%%%%%%%%%%%%%%%%%%%%%%%%%%%%%%%%%%%%%%
\begin{figure}[htbp]
	\begin{center}		
      {\includegraphics[width=0.45\textwidth]{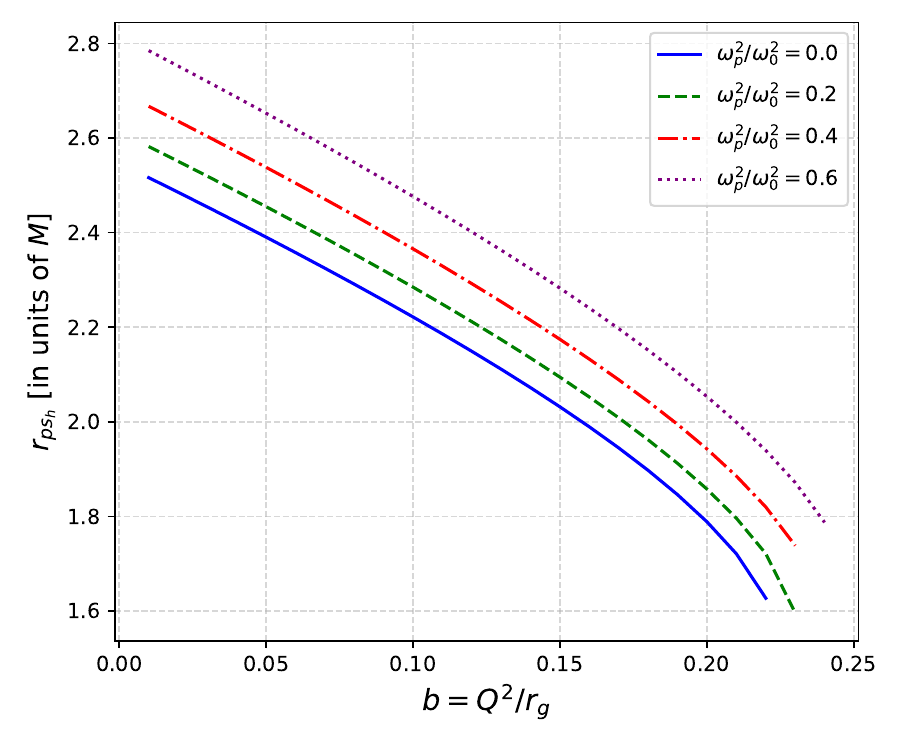}}
       {\includegraphics[width=0.45\textwidth]{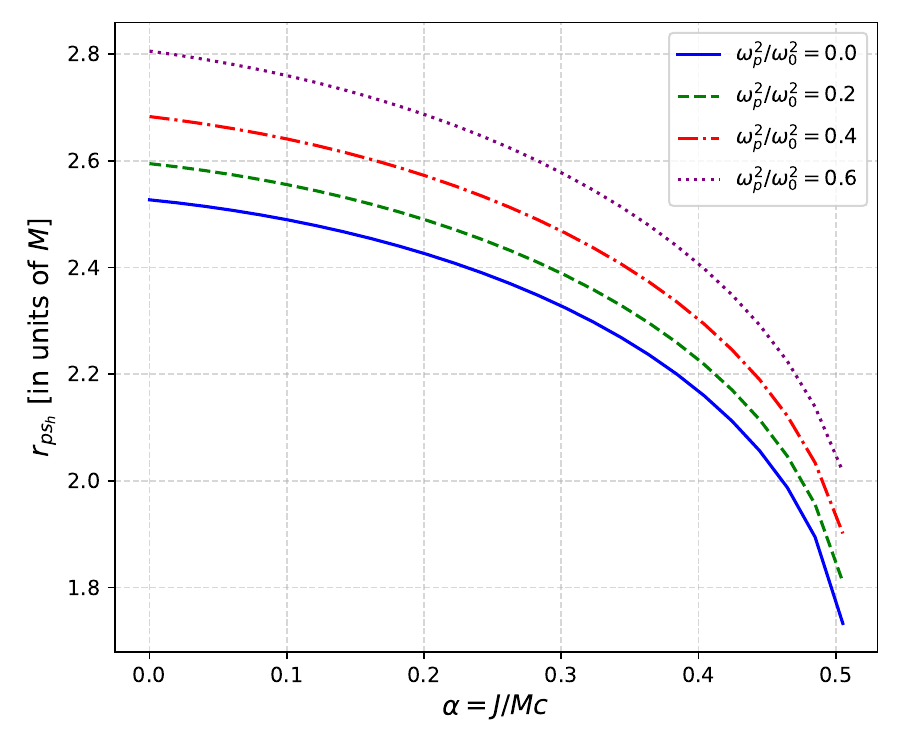}}
	\end{center}
	\caption{Photon sphere radius variation in the presence of a homogeneous plasma. (Top panel) Dependence on the charge parameter $b$ for $M=1$ and $\alpha=0.5$. (Bottom panel) Dependence on the rotation parameter $\alpha$ for $M=1$ and $b=0.2$.}
    \label{fig:HPPS1}
\end{figure}
%%%%%%%%%%%%%%%%%%%%%%%%%%%%%%%%%%%%%%%%%%%%%%%%%%
Fig.~\ref{fig:HPPS1} presents the variation of the photon sphere radius, $r_{\rm ph}$, in the presence of a homogeneous plasma. The upper panel shows its dependence on the charge parameter $b$ for fixed value of rotation parameter $\alpha = 0.5$. It is observed that increasing charge decreases the photon sphere radius, and this effect is further modulated by the plasma frequency ratio $\omega_{p}^2/\omega_{0}^2$, with higher plasma contributions shifting the curves upward. The lower panel illustrates the effect of the rotation parameter $\alpha$ for fixed  values of charge parameter $b = 0.2$. Similar to the charge case, an increase in spin leads to a reduction of the photon sphere radius, while stronger plasma effects enlarge it. Together, these results demonstrate the competing influences of homogeneous plasma, charge, and rotation: while charge and rotation act to shrink the photon sphere, the plasma tends to expand it, highlighting the nontrivial interplay between matter fields and black hole parameters in determining photon trajectories.

\subsection{\textbf{Circular light orbits in non-homogeneous plasma medium}}
Following a similar procedure as in the previous subsection, we consider the effective metric function for the non-homogeneous plasma medium given in Eq.~\ref{HrnonhomoPlas}. Differentiating Eq.~\ref{HrnonhomoPlas} with respect to the radial coordinate $r$ and applying the photon sphere condition, the governing equation for the photon sphere in the presence of a non-homogeneous plasma medium is obtained as
\begin{multline}
\frac{2 r_{ps}}{A(r_{ps})} \biggl[ 1 - A(r_{ps}) \frac{\eta_{0}}{\omega_{0}^{2} r_{ps}^{\nu}} \biggr]
- \frac{r_{ps}^{2} A'(r_{ps})}{A^{2}(r_{ps})} \biggl[ 1 - A(r_{ps}) \frac{\eta_{0}}{\omega_{0}^{2} r_{ps}^{\nu}} \biggr] \\
- \frac{r_{ps}^{2}}{A(r_{ps})} \cdot \frac{\eta_{0}}{\omega_{0}^{2}} 
\biggl[ \frac{A'(r_{ps})}{r_{ps}^{\nu}} - \frac{\nu A(r_{ps})}{r_{ps}^{\nu+1}} \biggr] = 0.
\end{multline}
Once again, owing to the analytical intractability of the equation, we employ numerical methods to obtain the solutions.
%%%%%%%%%%%%%%%%%%%%%%%%%%%%%%%%%%%%%%%%%%%%%%%%%%
\begin{figure}[htbp]
	\begin{center}		
      {\includegraphics[width=0.45\textwidth]{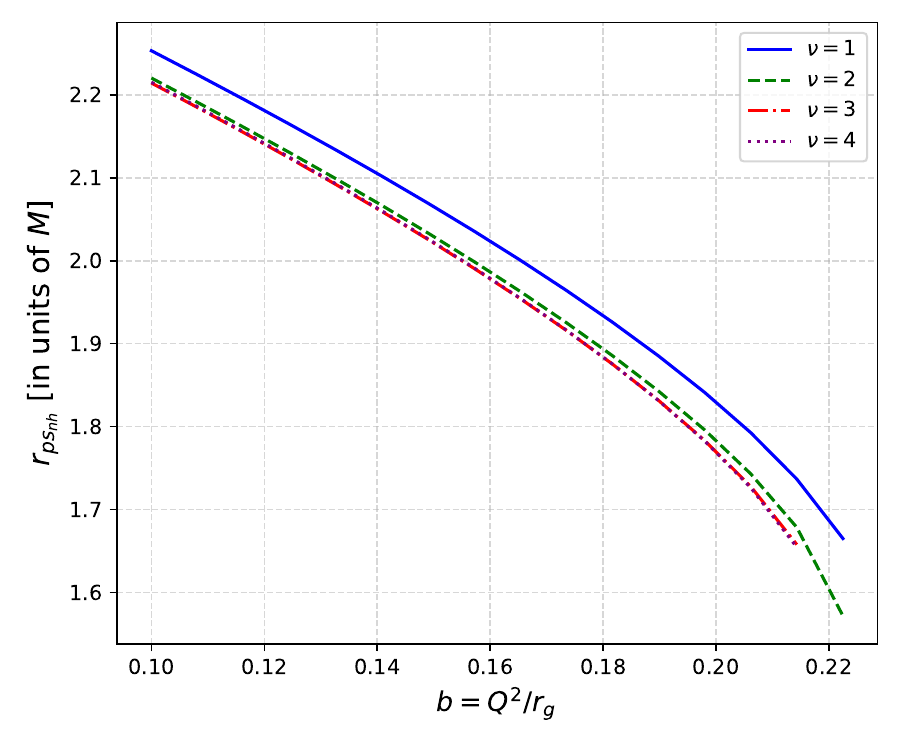}}
       {\includegraphics[width=0.45\textwidth]{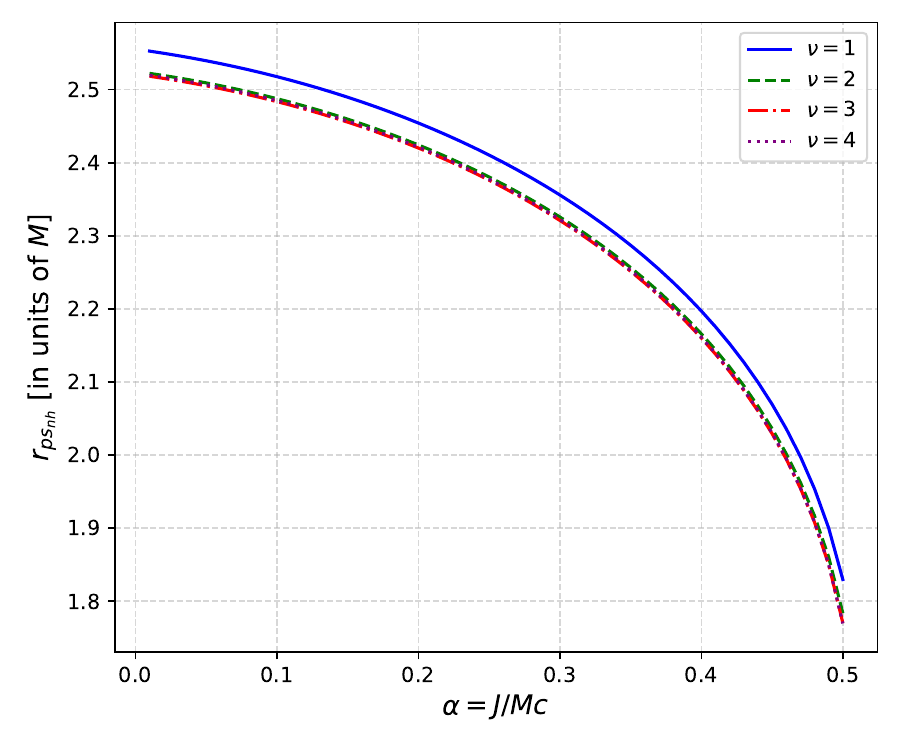}}
	\end{center}
	\caption{Photon sphere radius variation in the presence of a non-homogeneous plasma. (Top panel) Dependence on the charge parameter $b$ for $M=1$, $\eta_{0}=0.5$ and $\alpha=0.5$. (Bottom panel) Dependence on the rotation parameter $\alpha$ for $M=1$, $\eta_{0}=0.5$ and $b=0.2$.}\label{fig:NonHPPS1}
\end{figure}
%%%%%%%%%%%%%%%%%%%%%%%%%%%%%%%%%%%%%%%%%%%%%%%%%%
Fig.~\ref{fig:NonHPPS1} illustrates the variation of the photon sphere radius, $r_{\rm ph}$, in the presence of a non-homogeneous plasma. The upper panel shows the dependence on the charge parameter $b$ for fixed values of $\eta_0 = 0.5$, and $\alpha = 0.5$. It is observed that increasing the charge parameter reduces the photon sphere radius, while higher values of the plasma index $\nu$ shift the curves downward, indicating that inhomogeneous plasma enhances the shrinking effect of charge on the photon sphere. Notably, the photon sphere radius for $\nu = 3$ and $\nu = 4$ shows almost the same behavior, suggesting that the non-homogeneous plasma effects saturate beyond this range, which is consistent with the behavior observed in the deflection angle. The lower panel presents the variation with the rotation parameter $\alpha $ for $\eta_0 = 0.5$, and $b = 0.2$. Similar to the charge case, an increase in rotation leads to a decrease in the photon sphere radius. Notably, as the non-homogeneous plasma index increases from $\nu = 1$ to $\nu = 2$, a significant reduction in the photon sphere radius is observed. However, a further increase in $\nu$ from 2 to 4 results in only a marginal decrease. In point of view of a physical perspective, this suggests that the influence of plasma inhomogeneity on the photon sphere becomes less effective beyond a certain threshold, indicating a saturation behavior in the modification of photon orbits. Overall, the figure highlights that in non-homogeneous plasma, the photon sphere radius is more sensitive to both charge and rotation, with the plasma index $\nu$ amplifying their influence until it reaches saturation limit. This reveals a strong interplay between plasma inhomogeneity and black hole parameters, leading to notable modifications in the structure of photon orbits compared to the homogeneous plasma case.

\section{Result and Discussion}
In this study, we analyzed the gravitational lensing of massless particles in the spacetime of a Kerr–Sen black hole embedded in a magnetized, cold, pressureless plasma environment. Both homogeneous and inhomogeneous plasma distributions were incorporated in order to capture more realistic astrophysical conditions. The principal results of this work can be summarized as follows:
\begin{itemize}
    \item The deflection angle in the homogeneous plasma case increases with increasing plasma frequency, indicating that a denser uniform plasma distribution enhances the bending of light, in close agreement with the pioneering result first reported by Bisnovatyi-Kogan~\textit{et al.} and with subsequent seminal works in the literature~\cite{Bisnovatyi-Kogan:2017kii,Benavides-Gallego:2018ufb,Turimov:2018ttf,Yan:2019etp}. This behavior suggests that plasma dispersion plays a significant role in modifying photon trajectories in strong gravitational fields. Furthermore, the deflection angle decreases with increasing values of the rotation and charge parameters, showing that frame-dragging effects and the repulsive contribution associated with the black hole charge act to weaken the overall gravitational lensing effect.
    \item In the case of a non-homogeneous plasma medium, the deflection angle exhibits physical behavior similar to that associated with the black hole parameters in the homogeneous plasma case. Notably, for different values of the non-homogeneous index parameter, the variation in the deflection angle is much smaller compared to that observed in the homogeneous plasma scenario. This suggests that spatial inhomogeneity in the plasma distribution produces a weaker modulation of photon trajectories than uniform plasma density, indicating that homogeneous plasma effects dominate the overall gravitational lensing characteristics. Moreover, the nearly indistinguishable deflection angles observed for $\nu = 3$ and $\nu = 4$ indicate a saturation effect, suggesting that beyond a specific threshold, further changes in the non-homogeneous plasma profile yield no variation in bending. This observation implies that within this parameter range, light propagation is primarily governed by the combined influence of gravitational and non-homogeneous plasma effects, with higher-order alterations in the non-homogeneous plasma density profile exerting minimal influence on the deflection angle.
    \item The photon sphere radius in a homogeneous plasma decreases with increasing black hole charge and rotation, whereas stronger plasma effects lead to an expansion of the photon sphere. This behavior can be attributed to the dispersive nature of plasma, which possesses a frequency-dependent refractive index that modifies photon propagation. Consequently, the location of the photon sphere around the KSBH is not determined solely by spacetime geometry but is also regulated by the plasma medium through its influence on photon frequency, leading to a nontrivial modification of circular photon orbits.
    \item In the case of a non-homogeneous plasma medium, the photon sphere exhibits physical behavior similar to that observed under variations of the charge and rotation parameters in a homogeneous plasma environment. However, the photon sphere radius decreases with increasing non-homogeneous plasma index parameter any saturated beyond a threshold limit. Notably, the presence of black hole rotation weakens the influence of plasma inhomogeneity, indicating that frame-dragging effects partially suppress the impact of spatial plasma variations on circular photon orbits.
    \item The obtained results clearly demonstrate the deviation in light deflection caused by the presence of plasma when compared with the weak-field deflection angle of a KSBH in vacuum. In the absence of plasma, these results are qualitatively consistent with those reported by Uniyal~\textit{et al.}~\cite{uniyal2018bending} using exact elliptical integration and by Roy~\textit{et al.}~\cite{Roy:2025hdw} employing the material medium approach.
    \item The results obtained clearly highlight the impact of different plasma environments on the lensing properties of charged rotating black holes. However, in this study, we considered only cold plasma; in realistic astrophysical scenarios, hot plasma is also expected to influence the lensing characteristics. Next-generation experiments could, in principle, probe these lensing signatures in dispersive media, offering new insights into the interplay between strong gravity and plasma effects~\cite{Johnson:2023ynn,Genzel:2024vou}.    
\end{itemize}

\section*{Acknowledgments} 
Authors, SR and SM express sincere and deep gratitude to the Department of Physics, NITA, for providing the necessary research environment to complete this work. 
The author, SK, sincerely acknowledges IMSc for providing exceptional research facilities and a conducive environment that facilitated his work as an Institute Postdoctoral Fellow. One of the authors, HN, would like to thank IUCAA, Pune, for the support under its associateship program, where a part of this work was done. The author H.N. acknowledges financial support from the Anusandhan National Research Foundation (ANRF), through the Science and Engineering Research Board (SERB) Core Research Grant (Grant No. CRG/2023/008980).

%\appendix
%\renewcommand\theequation{\Alph{section}\arabic{equation}} % Only change equation numbering
%\setcounter{section}{1} % manually start from section A = 1
%\setcounter{equation}{0}
%\section*{Appendix A: }

%\bibliographystyle{IEEEtra}
\bibliographystyle{spphys}
\bibliography{KerrSen}
\end{document}